\documentclass[fleqn,usenatbib]{mnras}
\usepackage{savesym}
\usepackage{amsmath}
\usepackage{color}
\usepackage{graphicx}
\savesymbol{iint}
\savesymbol{iiint}
\usepackage{txfonts}
\restoresymbol{TXF}{iint}
\restoresymbol{TXF}{iiint}

\usepackage{times}

\newcommand{\Msun}{\ensuremath{\textrm{M}_{\odot}}}

\newcommand{\kms}{km\hspace{0.25em}s$^{-1}$}
\newcommand{\ergs}{erg\hspace{0.25em}s$^{-1}$}

\newcommand{\NaI}{\mbox{Na\hspace{0.25em}{\sc i}}}

\newcommand{\SII}{\mbox{S\hspace{0.25em}{\sc ii}}}

\newcommand{\SiII}{\mbox{Si\hspace{0.25em}{\sc ii}}}

\newcommand{\CaII}{\mbox{Ca\hspace{0.25em}{\sc ii}}}
\newcommand{\TiII}{\mbox{Ti\hspace{0.25em}{\sc ii}}}

\newcommand{\FeII}{\mbox{Fe\hspace{0.25em}{\sc ii}}}
\newcommand{\FeIII}{\mbox{Fe\hspace{0.25em}{\sc iii}}}
\newcommand{\CoII}{\mbox{Co\hspace{0.25em}{\sc ii}}}
\newcommand{\CoIII}{\mbox{Co\hspace{0.25em}{\sc iii}}}
\newcommand{\NiII}{\mbox{Ni\hspace{0.25em}{\sc ii}}}
\newcommand{\Fefs}{$^{56}$Fe}

\newcommand{\Cofs}{$^{56}$Co}
\newcommand{\Nifs}{$^{56}$Ni}

\newcommand{\KE}{\ensuremath{E_{\rm K}}}
\newcommand{\Dm}{\ensuremath{\Delta m_{15}(B)}}

\newcommand{\eg}{e.g.,\ }
\newcommand{\ie}{i.e.,\ }

\def\gsim{\mathrel{\rlap{\lower 4pt \hbox{\hskip 1pt $\sim$}}\raise 1pt \hbox {$>$}}}
\def\lsim{\mathrel{\rlap{\lower 4pt \hbox{\hskip 1pt $\sim$}}\raise 1pt \hbox {$<$}}}

\definecolor{myorange}{rgb}{0.9,0.6,0.0}

\definecolor{myredder}{rgb}{1.0,0.0,0.0}

\definecolor{mygreen}{rgb}{0.0,0.7,0.0}

\definecolor{myblue}{rgb}{0.0,0.0,1.0}

\definecolor{myyellow}{rgb}{0.92,0.89,0.00}

\begin{document}

\title[Nebular spectra of the SNe\,Ia 2007on and 2011iv]
  {The nebular spectra of the transitional Type Ia Supernovae 2007on and 
  2011iv: broad, multiple components indicate aspherical explosion cores}
\author[P. A.~Mazzali et al.]{P. A. Mazzali$^{1,2}$
\thanks{E-mail: P.Mazzali@ljmu.ac.uk}, 
C. Ashall$^{1,3}$, E. Pian$^{4}$,
M. D. Stritzinger$^5$, C. Gall$^{5,6}$, M. M. Phillips$^7$, 
\newauthor P. H\"oflich$^3$, E. Hsiao$^3$
\\
\\
  $^1$Astrophysics Research Institute, Liverpool John Moores University, IC2, Liverpool Science Park, 146 Brownlow Hill, Liverpool L3 5RF, UK\\
  $^2$Max-Planck-Institut für Astrophysik, Karl-Schwarzschild Str. 1, D-85748 Garching, Germany\\
  $^3$Florida State University, Tallahassee, USA \\
  $^4$IASF-Bo, Bologna, Italy\\
  $^5$Department of Physics and Astronomy, Aarhus University, Ny Munkegade 120,
DK-8000 Aarhus C,Denmark\\
  $^6$Dark Cosmology Centre, Niels Bohr Institute, University of Copenhagen, Copenhagen, Denmark\\
  $^7$Carnegie Observatories, Las Campanas Observatory, 601 Casilla, La Serena, Chile\\
}

\date{Accepted ... Received ...; in original form ...}
\pubyear{2017}
\volume{}
\pagerange{}

\maketitle

\begin{abstract} 
The nebular-epoch spectrum of the rapidly declining, ``transitional'' type Ia 
supernova (SN) 2007on showed double emission peaks, which have been interpreted 
as indicating that the SN was the result of the direct collision of two white 
dwarfs.
The spectrum can be reproduced using two distinct emission components, one
red-shifted and one blue-shifted. These components are similar in mass but have
slightly different degrees of ionization. They recede from one another at a
line-of-sight speed larger than the sum of the combined expansion velocities of
their emitting cores, thereby acting as two independent nebulae. 
While this configuration appears to be consistent with the scenario of two 
white dwarfs colliding, it may also indicate an off-centre delayed detonation 
explosion of a near Chandrasekhar-mass white dwarf.   
In either case, broad emission line widths and a rapidly evolving light curve 
can be expected for the bolometric luminosity of the SN.  
This is the case for both SNe 2007on and 2011iv, also a transitional SN\,Ia 
which exploded in the same elliptical galaxy, NGC 1404. 
Although SN\,2011iv does not show double-peaked emission line profiles, the 
width of its emission lines is such that a two-component model yields somewhat 
better results than a single-component model. Most of the mass ejected is in 
one component, however, which suggests that SN\,2011iv was the result of the 
off-centre ignition of a Chandrasekhar-mass white dwarf. 
\end{abstract}


\section{Introduction}
\label{sec:introduction}
 
Type Ia {\bf s}upernovae (SNe) are thermonuclear explosions of white dwarfs
(WD). The majority of these produce luminous transients, and are compatible with
the explosion of carbon-oxygen (CO) WDs which approach the Chandrasekhar limit
\citep{zorro}. Details of the accretion and explosion process remain highly
controversial, and it is also not clear whether all ``normal'' SNe\,Ia (those
which are normally used for cosmology) are the result of a single mechanism.  

Various less common sub-types of SNe\,Ia also exist. It is often debated whether
they are part of the same mechanism or the manifestation of different ones. 
Several scenarios have been developed that are thought to lead to the
thermonuclear explosion of a WD in a single-degenerate (SD) system or of both
WDs in a double-degenerate (DD) system. SD scenarios involve a CO-WD accreting H
or He from a non-degenerate companion
\citep{whelaniben,nomotokondo,livne90,livneglasner,livnearnett,woosleyweaver}.
DD scenarios envision two WDs with combined mass exceeding the Chandrasekhar
limit merging after their orbits shrink via emission of Gravitational Waves
\citep{ibentutukov,webbink84}. Although this should lead to the collapse of the
resulting massive ONeMg WD \citep{nomoto82}, it may result in a thermonuclear 
explosion if the merger is a ``violent'' one \citep{pakmor10}. In an extreme
case  the WDs may suffer a direct collision and explode by compression. This is
expected  to happen mostly in dense stellar environments such as globular
clusters  \citep{rosswog09}, but it may be aided if their orbits are focused by
the presence  of a third or fourth system member \citep{kushnir13,fang2017}. The
details of the explosion depend on the masses of the WDs involved, but the
morphology of a DD explosion is always expected to be more aspherical than that
of a SD explosion. 
If the collision has non-zero impact parameter, the explosion
may lead to a two-component ejecta structure \citep{dong15}.

While luminous SN\,1991T-like SNe \citep{filipp91T,phillips91T} may just be a
``more efficient" extension of normal SNe\,Ia \citep{m95,sasdelli14,zhang16}, at
the low end of the SN\,Ia luminosity distribution different mechanisms may be
operating.  SN\,1991bg \citep{filipp91bg,leib91bg}, the prototypical ``fast
decliner", and the class which takes its name, which is characterised by spectra
showing lines of low-excitation ions \citep{mazzali97}, have been suggested to
be the product of a merger of two WDs below the Chandrasekhar mass
\citep{mh12,pakmor2011}. 

SNe that straddle the 1991bg class and the group of normal SNe\,Ia are known as
``transitional''. They have rapidly evolving light curves, but spectra that are
not as peculiar as those of SN\,1991bg. Just like the 1991bg class, they tend to
occur in non-star-forming galaxies \citep{ashall2016a}. SN\,1986G, which was the
first SN\,Ia observed to show the \TiII\ absorption complex near 4500\,\AA\
which would later become the hallmark of the sub-luminous 1991bg class
\citep{phillips87,cristiani92}, can be explained by a low-energy explosion that
still has a mass compatible with the Chandrasekhar mass \citep{ashall2016a}.
SN\,2003hv \citep{leloudas03hv}, which was singled out for its flat-topped
emission line profiles \citep{motohara06} and the noticeable shift of some
emission lines \citep{maeda10}, was suggested to be a sub-Chandrasekhar-mass
explosion based on the behaviour of its light curve and nebular spectra
\citep{mazzali03hv}.  Finally, based on its apparently double-peaked
emission-line profiles, SN\,2007on was suggested to be an example of a head-on
collision of two white dwarfs \citep{dong15}. 

Most of these suggestions have been based on the study of nebular emission
lines. Nebular spectra are very useful in SN studies as they reveal the
properties of the inner ejecta, which are hidden under a blanket of highly
opaque material at earlier times. 

\begin{figure}
 \includegraphics[width=88mm]{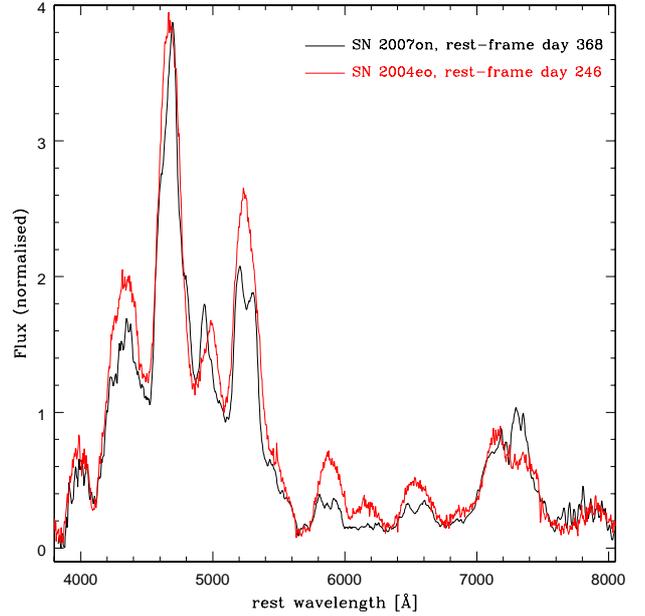}
\caption{A nebular spectrum of SN\,2007on (black) compared to one of SN\,2004eo
(red). The spectra have been normalised at the peak of the strongest emission 
line.}
\label{fig:nebcomp07on_04eo}
\end{figure}

A late-time spectrum of SN\,2007on \citep{gall2018} is shown in Fig.
\ref{fig:nebcomp07on_04eo}, where it is compared to a nebular spectrum of the 
normal, but rapidly declining SN\,Ia 2004eo \citep{pastorello04eo}. The figure 
highlights that the two spectra are similar, but SN\,2007on, apart from having 
slightly narrower emission lines, is characterised by double peaks in several
emission lines as well as by a shift in the position of some of the emission
peaks. This highlights why SN\,2007on stands out in the nebular phase, although
it is not particularly peculiar at earlier times \citep{ashall2018}.

Here the nebular spectrum of SN\,2007on is analysed in order to determine the
properties of the explosion (Section \ref{sec:models}). Section \ref{sec:disc}
contains a discussion of SN\,2007on, and Section \ref{sec:11iv} an analysis of
SN\,2011iv, a transitional SN that exploded in the same host galaxy as SN
2007on, conducted with the same criteria. The nebular data used in this paper
were presented by \citet{gall2018}.

\section{Synthetic Models for SN\,2007on}
\label{sec:models}

The nebular spectrum was modelled using our non-local thermodynamic equilibrium
(NLTE) code. The code has been applied to SNe of different types, \eg SNe\,Ia
\citep{mazzali04eo}, SNe\,Ib/c \citep{mazzali10}. Details can be found in those
and other papers. Briefly, the energy deposited in the SN nebula by the
particles emitted by the radioactive decay of \Nifs\ into \Cofs, and hence into
\Fefs, is used to heat the SN nebula via collisions. Cooling via (mostly
forbidden) line emission balances the heating and gives rise to the observed
spectrum.  In the case at hand we want to investigate the morphology of the SN
ejecta.  Therefore, instead of using a stratified density distribution and an
explosion model as we did for example in \citet{mazzali15}, we use a simple
1-zone version of the code, which is designed to bring out the most important
characteristics of a spectrum and the mass of \Nifs\ to a good approximation,
but cannot be used to estimate the total ejecta mass with great accuracy.  This
method is however adequate here as a first step to determine the properties of
the inner ejecta, which is our main interest in the case of SN\,2007on. 
The discussion below follows the process of investigation, in an attempt to show
how we arrived at the solution which is proposed here. 

We begin by establishing the basic parameters for the model. The spectrum has
been corrected for the recession velocity of the host galaxy (1947\,\kms), with
a correction for the specific velocity at the location of the SN ($-95$\,\kms)
\citep{franx89}, resulting in a recession velocity of 1852\,\kms. A distance
modulus to the galaxy $\mu = 31.20$ is used, and reddening is assumed to be
negligible \citep{gall2018}. The spectrum was obtained 353 days after $B$-band
maximum. We used a risetime of 17.4 days, as estimated by \citet{gall2018}. 
This is consistent with rapidly evolving, sub-luminous SN\,Ia 
\citep{heyden10,conley06,Ganeshalingam2011,mazzali03hv,hsiao15}. We therefore
use a rest-frame epoch $t = 368$ days.  

\begin{figure*}
 \includegraphics[width=139mm]{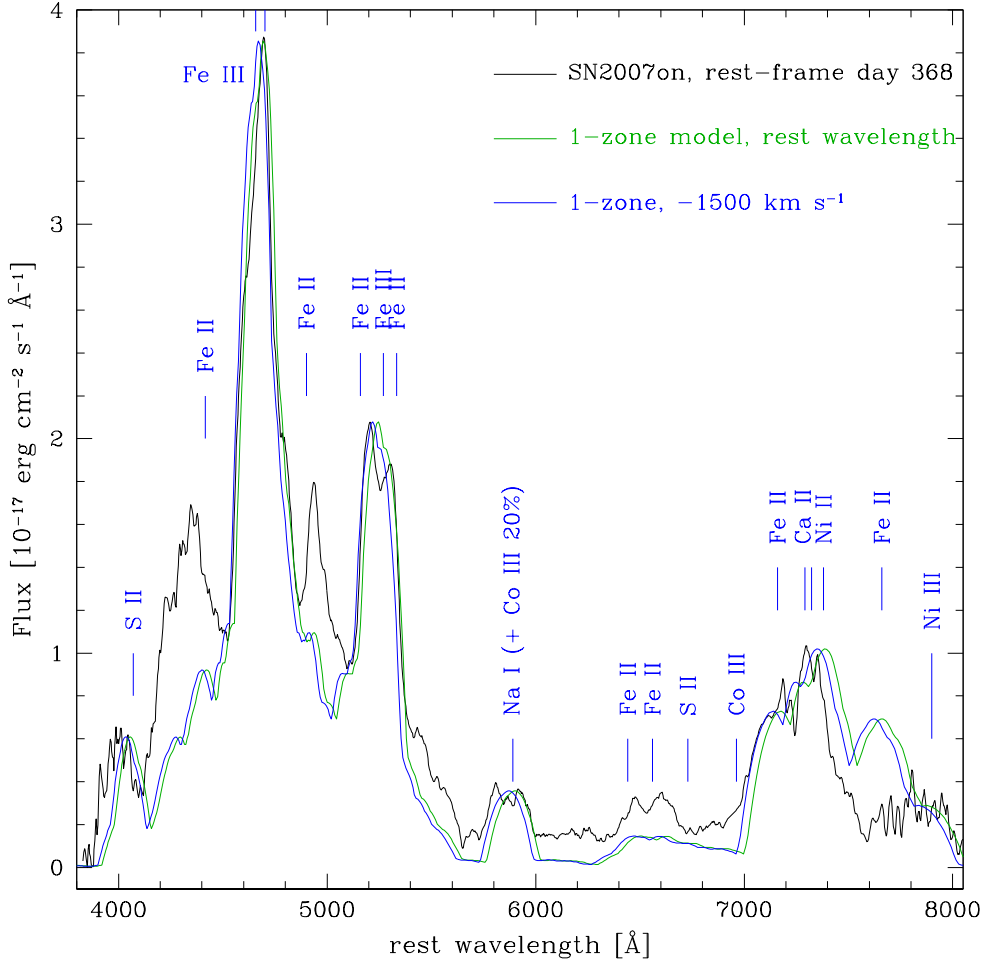}
\caption{The nebular spectrum of SN\,2007on (black) and a 1-zone, broad-line 
model, shown both at rest (green) and with a blueshift of 1500\,\kms\ (blue).}
\label{fig:sn2007on_1comp_shift1}
\end{figure*}

Our first attempt was to compute a simple 1-zone model to match the entire
spectrum. The primary criterion we use to judge our models is their ability to
match both the [\FeIII]-dominated emission line near 4800\,\AA\ and the
[\FeII]-dominated emission line near 5200\,\AA. This is because these are the
strongest lines, so that reproducing them should yield the most realistic
estimate of the mass of \Nifs. Additionally, because these two lines are
dominated by different ionization stages of the same element, reproducing their
ratio gives a reasonable estimate of the temperature/density conditions in the
SN nebula. The model we computed is shown in Figure
\ref{fig:sn2007on_1comp_shift1} as a fully drawn (green) line.  It has a
boundary velocity of 6500\,\kms\ and it includes 0.19\,\Msun\ of \Nifs\ and a
total mass of 0.27\,\Msun\ inside that velocity. We call this the
``broad-line model". The \Nifs\ mass is similar to that obtained by
\citet{gall2018} using a simple approximation (0.25\,\Msun). The nebular line
velocity is small, as expected for a rapidly declining SN\,Ia
\citep{mazzali98,zorro}, but it is large for the decline rate of SN\,2007on
($\Delta {m}_{15}(B)$ = 1.96\,mag, similar to SN\,1991bg). We note that
\citet{burns2014} showed that $\Delta {m}_{15}(B)$ is not a good discriminant at
these very fast decline rates. \citet{gall2018} show that these transitional SNe
are better ordered when plotted against the parameter sBV.

As Fig. \ref{fig:sn2007on_1comp_shift1} shows, a 1-zone model does not provide a
very good fit. The problem is not so much the relative strength of the [\FeII]
and [\FeIII]-dominated features -- this can be reproduced using a small but not
negligible mass of stable Fe ($\approx 0.05$\,\Msun) and stable Ni ($\approx
0.02$\,\Msun), which cool the gas, leading to a reduced ionization and
suppressed [\FeIII] emission -- but rather the shape of the various lines, the
strongest of which are labelled in the figure. The strongest [\FeIII] line near
4700\,\AA\ is narrower than the synthetic emission, especially near the peak,
but the composite [\FeII]-[\FeIII] emission near 5200\,\AA\ is broader and bluer
than the model line and shows multiple peaks. Line blending and overlap are
automatically taken into account by our code. Similar problems are seen in
weaker lines: the Na/Co line near 5900\,\AA\  (which is dominated by \NaI\,D) is
only reproduced on the red side, while the observed split is clearly not
reproduced; the Fe/Ca/Ni structure near 7200\,\AA\ is also only approximately
reproduced, {\em although it may be affected by the data reduction process: the
sharp absorptions do not seem to be natural for a SN}. Other lines are
reproduced in wavelength but not in strength. This affects in particular a
number of [\FeII]-dominated features (4400, 4900, 6500, 6600\,\AA). A strong
[\FeII] emission is predicted near 7600\,\AA\ but not observed with a comparable
strength. For this and other weak [\FeII] lines the problem may be inaccurate
collision strengths.  A [\SII] line appears to reproduce - at least partially -
a line near 4000\,\AA. 

\begin{figure*} \includegraphics[width=139mm]{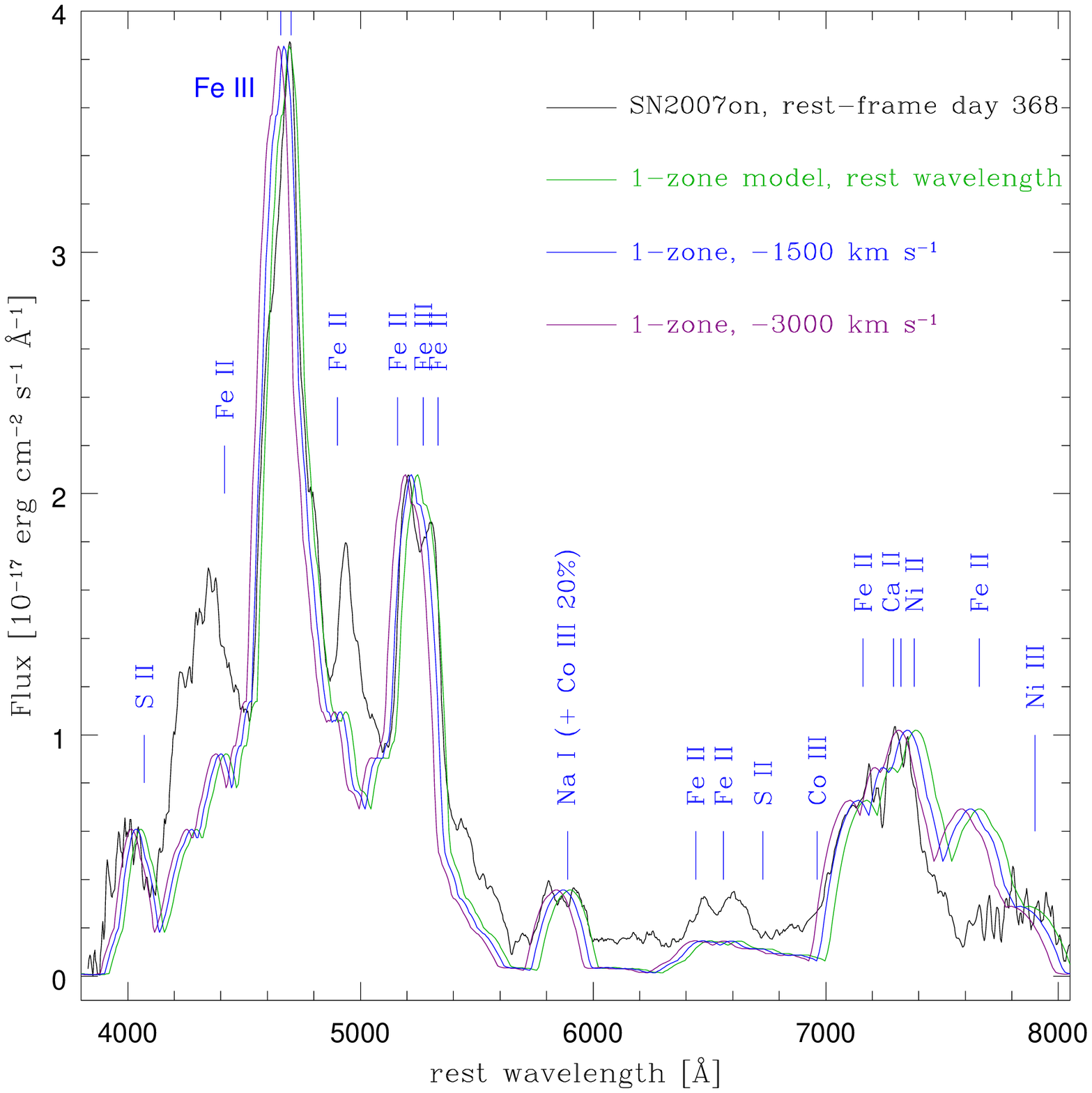}
\caption{Same as Figure \ref{fig:sn2007on_1comp_shift1}, but now also showing a
blueshift of 3000\,\kms\ (violet line).}
\label{fig:sn2007on_1comp_shift2}
\end{figure*}

For most SNe\,Ia, a 1-zone model is a good starting approximation, which can
then be improved using a stratified model to obtain for example a better
description of the line profiles. In the case of SN\,2007on, however, the
discrepancies between the observed spectrum and the 1-zone model are too complex
to be resolved with a stratified 1-D model.  This confirms the suggestion of
\citet{dong15} that multiple components are required. An alternative option may
be that the double peaks are caused by self-absorption, but it seems unlikely
that SN\,2007on is the only SN\,Ia for which this occurs. Also, not all
emissions are affected (and some features are the result of the blend of many
lines). SN\'1986G showed double peaks on the \NaI D line, which was caused by
interstellar \NaI D absorption in the host galaxy. SN\,1986G was heavily
reddened, but this is not the case for SN\,2007on. 

In order to figure out how the nebular spectrum of SN\,2007on is formed, we
began by trying to match the position of the observed lines. First we shifted
the 1-zone synthetic spectrum bluewards by 1500\,\kms. This is shown in Fig.
\ref{fig:sn2007on_1comp_shift1} as a blue line. The blue-shifted model now
matches the blue side of the 5200\,\AA\ emission, including the narrow peak. It
also matches the blue side of the emissions at 4700\,\AA\ and 7000-7400\,\AA,
but it is too blue for the 4700\,\AA\ peak and not blue enough for the emission
at 5900\,\AA.  It matches the [\SII] emission near 4000\,\AA\ better, while
other weaker lines do not change significantly. 

In order to match the blue side of the \NaI\,D/[\CoIII] line a larger blueshift
is required. The line coloured in violet in Fig. \ref{fig:sn2007on_1comp_shift2}
shows the 1-zone synthetic spectrum with a blueshift of 3000\,\kms. With this
larger blueshift, emission lines of lighter elements match the data better. 
This is not just the case for the \NaI\,D/[\CoIII] line but also for the [\SII]
line near 4000\,\AA\ and the [\CaII] line near 7300\,\AA, although the
contribution of [\NiII] and the possibly spurious narrow absorptions make any
conclusions about that feature uncertain. However, all synthetic Fe lines are
now too blue. 

So, while a multi-component model seems to be required to explain the spectrum,
its details are not immediately inferred just from inspection.  It is necessary
to decompose the observed spectrum into different components. These may be
characterised by narrower emission lines, of width $\sim 4000$\,\kms, matching
the width of the narrow emission peaks.

\begin{figure*} 
\includegraphics[width=139mm]{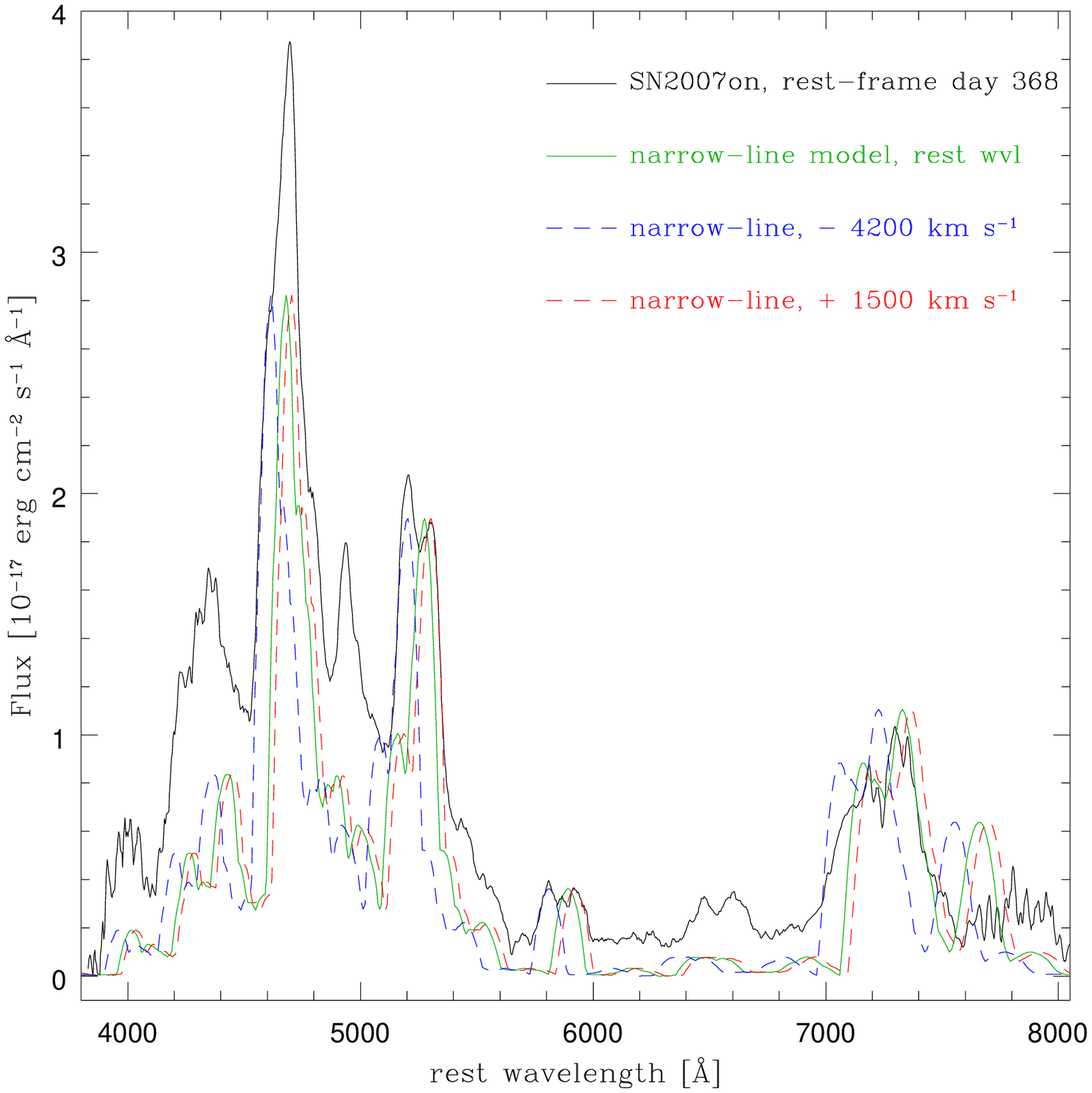}
\caption{The nebular spectrum of SN\,2007on (black) and a narrow-line model 
shown both at rest (green) and with a blueshift of 4200\,\kms\ (dashed, blue line) 
and a redshift of 1500\,\kms\ (dashed, red line) to match narrow features.}
\label{fig:SN2007on_d353_1narrowlinemodel_rest_blue_redshift}
\end{figure*}

One place to start is to notice that the peaks that are clearly split (\eg those
at 5200 and 5900\,\AA) show much narrower components. An attempt to fit these is
shown in Fig. \ref{fig:SN2007on_d353_1narrowlinemodel_rest_blue_redshift}, where
we use a single, ``narrow-line" 1-zone model. This narrow-line synthetic
spectrum is characterised by a width of only 4200\,\kms, matching the width of
the narrow emission peaks. This narrow-line synthetic spectrum matches the
position of the peaks near 4700 and 7200-7300\,\AA, but it fails to match either
peak of the lines near 5200 and 5900\,\AA. Additionally, a single narrow-line
model is unable to match the high peak ratio of the 4800\,\AA\ v. 5200\,\AA\
emissions.  This is because the model has rather high density and consequently a
low degree of ionization, which favours the \FeII-dominated 5200\,\AA\ emission.
Only removing elements that cool the gas can cause the ionization to increase,
so for example the [\SII] emission is poorly reproduced. This means that a
simple shift will not work and multiple components with different properties and
different shifts in wavelength are needed. We therefore proceed to build a
2-component model by combining two separate 1-zone models. 

We are guided in the construction of a 2-component model by the relative shift
of the unblended emission peaks. The position of the two emission peaks of the
\NaI\,D/[\CoIII] line indicates an offset of about 6000\,\kms. Figure
\ref{fig:SN2007on_d353_1narrowlinemodel_rest_blue_redshift} shows the
narrow-line 1-zone model shifted to the red and the blue, respectively, in order
to match the emission peaks of the lines near 5200 and 5900\,\AA. The red dashed
line has a redshift of 1500\,\kms. With this redshift the synthetic spectrum
matches in position and width the peaks at 5300 and 5900\,\AA, which are the red
components of the emission features near 5200 and 5850\,\AA, respectively.  The
blue dashed line has a blueshift of 4200\,\kms, which matches the bluer peak of
the same features. Interestingly, it also matches the blue side of the strong
4700\,\AA\ emission as well as the position of the peak near 4350\,\AA,
confirming that two components with different line-of-sight velocity may be
needed to reproduce the observations.

\begin{figure*}
\includegraphics[width=139mm]{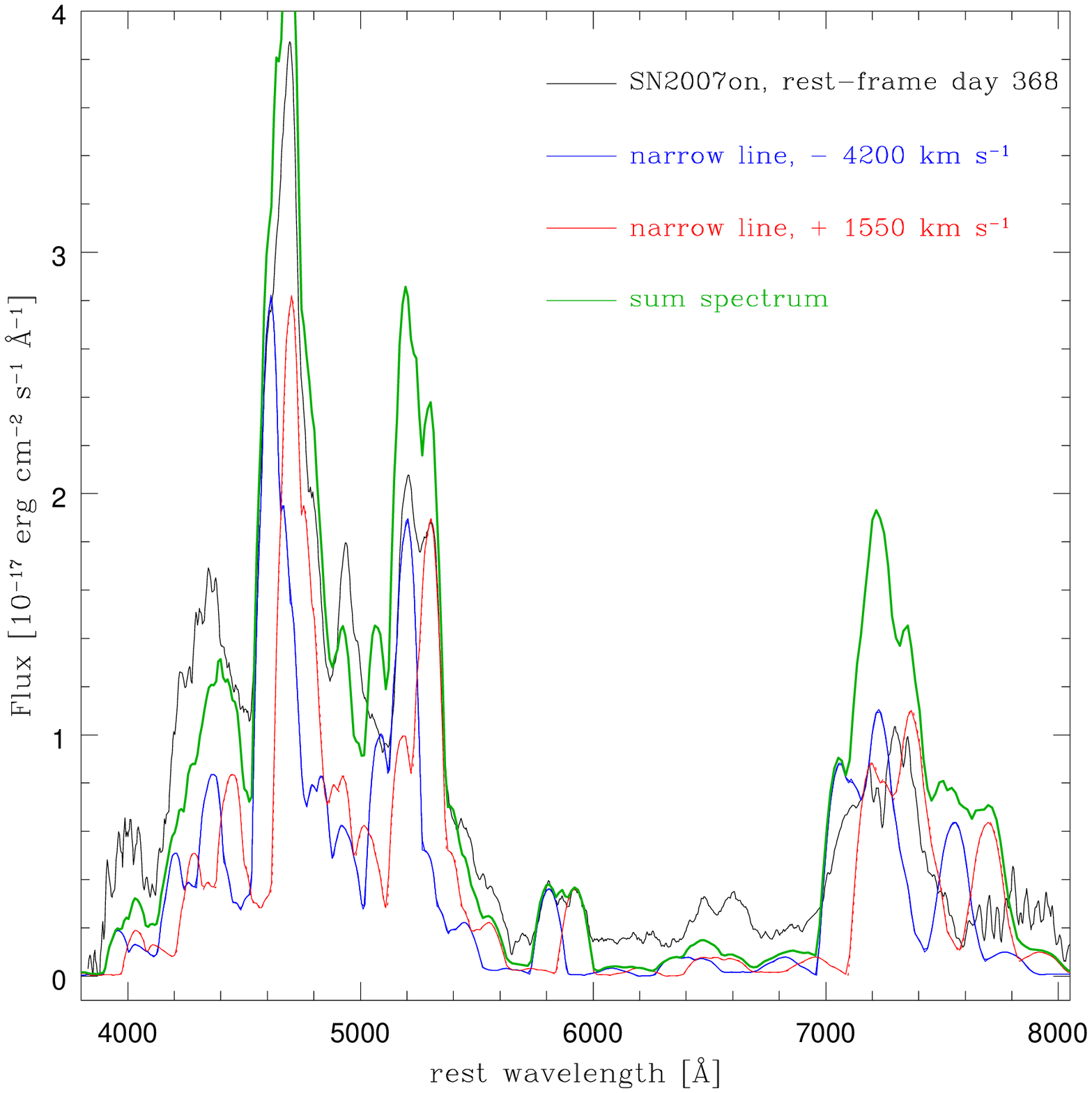}
\caption{Same as Figure
\ref{fig:SN2007on_d353_1narrowlinemodel_rest_blue_redshift}, but now the green 
line is the sum of the two spectra model.}
\label{fig:SN2007on_d353_1narrowlinemodel_blue_redshift_sum}
\end{figure*}

Although shifting the narrow-line 1-zone spectrum in wavelength can match the
position of different emission peaks, simply summing the blue- and red-shifted
1-zone spectra to compute a combined spectrum does not match the observations
(Fig. \ref{fig:SN2007on_d353_1narrowlinemodel_blue_redshift_sum}), although it
works amazingly well in places (\eg the [\FeII]-dominated emission near
5200\,\AA, the \NaI\,D/[\CoIII] line near 5900\,\AA). Yet, this gives us an
indication of the kind of procedure we must follow to obtain a fit with a
2-component model.  

To begin with, the two components cannot be as luminous as what was shown in
Figures \ref{fig:SN2007on_d353_1narrowlinemodel_rest_blue_redshift} and 
\ref{fig:SN2007on_d353_1narrowlinemodel_blue_redshift_sum}. Secondly, more
emission should be achieved in some of the features that are under-reproduced.
This includes [\SII] near 4000\,\AA\ and several [\FeII]-dominated lines.
Conversely, the [\FeII]-dominated emission near 7600\,\AA\ should be essentially
absent. While increasing the flux of lines of intermediate-mass elements such as
S is possible at the cost of a lower temperature, changing the ratios of \FeII\
transitions is not something we can manufacture, as it depends on the atomic
data that are available. If these are not accurate or complete, we can expect
discrepancies between the 2-component model and the data. 

We built two distinct 1-zone models to represent the two different components to
the spectrum. The models are characterised by different degrees of ionization.
The velocity width of the individual components as well as their separation were
taken as derived above for the narrow-line 1-zone model shown in Fig.
\ref{fig:SN2007on_d353_1narrowlinemodel_rest_blue_redshift}.  We
computed the two 1-zone models iteratively, in an effort to obtain the ``best''
solution, but did not attempt to match every detail. Our aim was to explore how
well the combination of 2 components can match the data and to determine the
basic physical properties of these components.   

Based on the indications above, we need to construct two narrow-line models,
each with width $\sim 4000$\,\kms, separated by $\sim 5500$\,\kms. One model is
slightly red-shifted ($\sim 1500$\,\kms), and the other significantly
blue-shifted ($\sim 4200$\,\kms). From Figure 
\ref{fig:SN2007on_d353_1narrowlinemodel_blue_redshift_sum} the blue-shifted
model is needed to reproduce the blue shoulder in the emission near 4700\,\AA\
as well as the bluer peak in the 5200\,\AA\ emission, and it should be very
similar to what we used in that figure. The red-shifted model, on the other
hand, needs to be less luminous, so that the red-side peaks are not over-fitted.

\begin{figure*}
 \includegraphics[width=139mm]{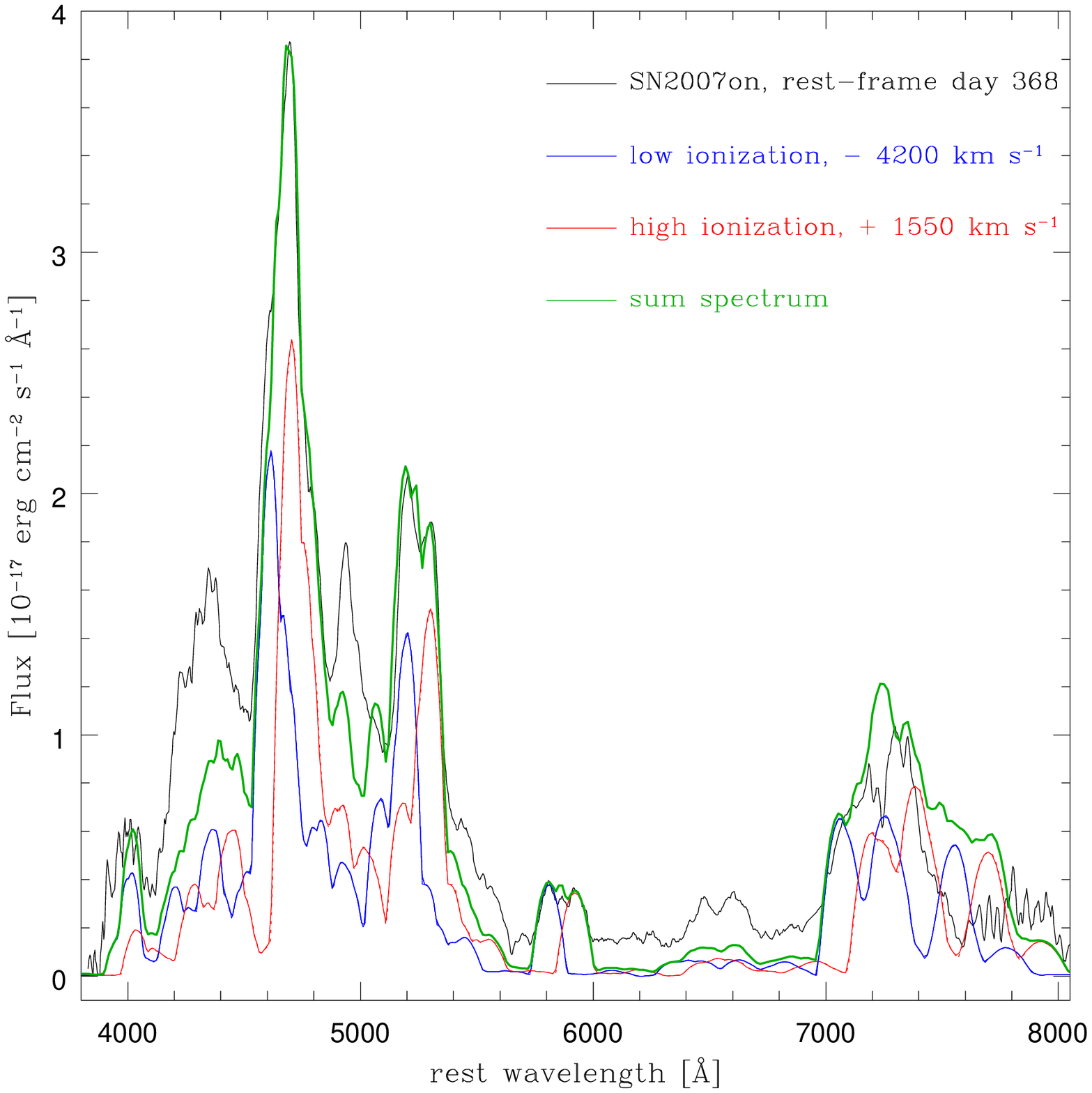}
\caption{The nebular spectrum of SN\,2007on (black) and the 2-component model 
(blue and red shifted 1-zone components are shown in blue and red, 
respectively) as discussed in the text. The green line is the sum of the two 
components.} 
\label{fig:SN2007on_2comp}
\end{figure*}

The best solution we identified is shown in Figure \ref{fig:SN2007on_2comp}. The
two component spectra are shown by the blue and red lines, respectively, while
the green line is the sum of the two (assuming no radiation transport effects). 
The combination of narrow emission line width and component velocity separation
yields a combined emission line width similar to that of the observed spectral
features. 

One spectrum is blue-shifted by 1550\,\kms. The nebula that emits this spectrum
has $0.084$\Msun\ of \Nifs\ included within the boundary velocity (4000\,\kms). 
Additionally, a small amount of stable Ni (0.003\,\Msun) is required to form the
7380\,\AA\ line (which appears near 7250\,\AA). The presence and strength of
this line are uncertain given the probably spurious absorption that affects the
data. If some stable Ni is synthesised in the explosion this is typically
thought to be at high density near the core of a massive (near-Chandrasekhar
mass) CO white dwarf.  Some stable Fe is also usually synthesised in this case
\citep[see, \eg][]{iwamoto99}. Therefore, we also add a small amount of the
latter (0.006\,\Msun). Stable Fe helps to increase the strength of the [\FeII]
lines but this is achieved at the cost of a reduced overall ionization balance
and therefore a reduced strength of the [\FeIII] emission, in particular
relative to [\FeII] lines. Intermediate-mass elements (IME) must also be
present, if the emission line near 4000\,\AA\ is blue-shifted [\SII]. A small
amount of S (0.002\,\Msun) is sufficient for this purpose. As Si and S are
usually synthesised together, with a ratio of $\sim 2$ -- 3 to 1, we included
0.005\,\Msun\ of Si. We also included small amounts ($10^{-4}$ to $10^{-3}$) of
Mg, Ca (useful in the emission complex near 7300\,\AA) and Na (required for the
emission at 5900\,\AA). This blue-shifted spectrum matches only part of the
strongest line near 4800\,\AA, it reproduces the bluer peak of the line at
5200\,\AA\ in position but not in flux, and it successfully matches the blue
part of the \NaI D/[\CoIII] emission. 

The total mass of this blue-shifted nebula, inside the outer boundary velocity
(4000\,\kms), is $\sim 0.10$\,\Msun.  The outer boundary of the nebula implies
that this is just the inner core of the SN ejecta, which is the only part that radiates nebular lines at late times.
Fe-group elements are the dominant constituents of this innermost region, as in
typical SNe\,Ia. The mass enclosed within the outer boundary velocity is
somewhat small for a SN\,Ia. The fairly rapidly declining, normal SN\,Ia 2004eo
had $\approx 0.23$\,\Msun\ inside this velocity \citep{mazzali04eo}. On the
other hand, sub-luminous SNe\,Ia that have been singled out as possibly being
the result of different mechanisms have significantly smaller masses: SN\,2003hv
had $\approx 0.05$\,\Msun\ inside the same velocity \citep{mazzali03hv}, and
SN\,1991bg had $\approx 0.08$\,\Msun\  \citep{mh12}. 

The second component spectrum in our 2-component model is red-shifted by
4200\,\kms. The nebula that emits this spectrum is quite similar in properties
to the one that emits the blue-shfted spectrum, but is characterised by a 
somewhat higher degree of ionization, which is required to match the stronger
red-shifted component of the [\FeIII] emission near 4700\,\AA. Within an outer
boundary velocity of 4200\,\kms, which is only marginally larger than the
velocity of the blue-shifted model, this component includes a similar \Nifs\
mass as the blue-shifted one ($0.084$\Msun). Unlike the blue-shifted model,
intermediate-mass elements in the red-shifted model are negligible, but the very
small content of Na and Ca (both $\sim 10^{-4}$\,\Msun) is sufficient to give
rise to emission in the intrinsically strongest lines. The low abundance of IMEs
is important for obtaining a high degree of ionization, such that the ratio of
the two strongest lines favours the [\FeIII]-dominated 4700\,\AA\ feature more
than in the blue-shifted model. The ionization state is controlled by the
balance of heating and cooling. Heating is provided by \Nifs\ decay, cooling is
provided by line emission. The most important factor influencing ionization is
density, and in particular electron density, which controls the recombination
rate (which is $\propto \rho^2$). The presence of species that only contribute
to cooling but not to heating adds free electrons to the gas and reduces the
average ionization because these electrons can recombine with other ions. Hence,
in order to keep the ionization high, it helps if the gas is mostly composed of
species that contribute to heating as well as cooling, \ie\ radioactive
isotopes. Removing stable Fe-group species would also be helpful, but we need to
consider the presence of the [\NiII] line. The stable Ni content is
0.004\,\Msun, although again establishing its mass from the emission line near
7300\,\AA\ is complicated by the shape of the spectrum. The mass of stable Fe is
0.007\,\Msun. With a total mass of $\approx 0.10\,\Msun$ and a slightly larger
outer boundary velocity, the densities in the red-shifted model are lower than
in the blue-shifted one, and the ionization degree higher. This leads -- as
desired -- to an \FeIII-dominated spectrum, as shown by the ratio of the two
strongest lines (red line in Fig. \ref{fig:SN2007on_2comp}).  This spectrum
matches only part of the strongest line near 4700\,\AA\ and the redder peak of
the line at 5200\,\AA, and it successfully matches the red component of the
\NaI\,D/[\CoIII] emission near 5900\,\AA. It also reproduces the approximate
position of the peaks of the \FeII/\NiII/\CaII\ complex at 7100-7400\,\AA. 

Similar considerations about the mass of the red-shifted component hold as for
the blue-shifted one (see above).  It is small for a normal SN\,Ia, but it is
larger than in both merger and sub-Chandrasekhar models. Neither model seems to
reproduce any previously modelled SN\,Ia \citep[\eg][]{mazzali04eo}.  

A combined spectrum was obtained by summing the two components, neglecting any
radiation transport between them. This is both because we assume the gas to be
optically thin and -- primarily, because we do not know the relative position of
the two components with respect to our line of sight. The combined spectrum is
shown as a green line in Fig. \ref{fig:SN2007on_2comp}. Most of the complex
emission profiles are well reproduced by this spectrum, in shape and wavelength
if not always in strength. We could highlight the tiny peak in the centre of the
emission line near 5890\,\AA, which is the result of the sum of two very similar
individual parabolic components, but at this low flux level for the observed
spectrum we cannot exclude that the agreement with what we see in the data is
just a coincidence. Interestingly, also some of the weaker emissions (\eg the
[\FeII]-dominated emissions near 4400\,\AA\ and 6500\,\AA\ and the
\FeII/\NiII/\CaII\ structure at 7200--7500\,\AA) are much better reproduced by
the combined model than by the individual components. This lends further
credence to a model for SN\,2007on where two separate components contribute to
the nebular spectrum. The intensity of several [\FeII] lines is still too weak, and this may have to do with inaccurate atomic data, in particular collision strengths.

Although each component has rather small mass, the combined mass of the two
components inside their respective outer boundaries exceeds 0.2\,\Msun. This is
only slightly smaller than but comparable to a Chandrasekhar-mass explosion with
normal energy ($\approx 0.25$\,\Msun).

\section{Discussion: what was SN\,2007on?} 
\label{sec:disc}

The nebular spectrum of SN\,2007on can be explained assuming it is the sum of
two individual components, with similar line width and slightly different
density, composition and ionization degree, receding from one another with a
projected line-of-sight velocity of 5750\,\kms. 

Could these two components be part of the same explosion? It is natural to
imagine that they originated at the same time, and this was our modelling
assumption. The two 1-zone nebulae that we have defined cannot be physically
co-located, as they differ in density, composition and ionization structure. The
fact that the boundary expansion velocities of the two nebulae are similar, 4000
and 4200\,\kms, respectively, but at the same time the line-of-sight velocity
separation is 5750\,\kms, suggests that it is possible that the two nebulae are
contiguous. The real velocity separation is likely to be larger than 5750\,\kms,
and could be similar to or larger than the sum of the expansion velocities of
the two emitting regions we have modelled, 8200\,\kms.  In this case these two
nebulae, which make the major contribution to the nebular emission, would be
separated as bulk, each nebula expanding essentially into free space,
independent of the other one. Interestingly, even if the two nebulae were
actually moving along the line of sight and their velocity of separation was
indeed 5750\,\kms, their respective innermost parts, the regions with combined
expansion velocities lower than the separation velocity (i.e. $v \sim
2800$\,\kms), would never come into contact. On the other hand, regions with
combined velocities exceeding the separation velocity would interact at some
point in the life of the SN.  The question then is: Where did the two nebulae
come from?

The region that contributes to emission in the nebular phase is only the
innermost part of the ejecta of SN\,2007on: there is material at higher
velocities, containing IME in a significant fraction, which gives rise to the
early-time spectra of the SN \citep{gall2018,ashall2018}. The mass ejected in
SN\,2007on is obviously larger than what is included in the individual nebular
models. Further information about the outer layers can be obtained from the
early phases of the evolution of the SN. In a parallel paper, \citet{ashall2018}
find that the early-time spectra of SN\,2007on may be explained by a
normal-to-low energy explosion of a Chandrasekhar-mass progenitor. 

A scenario that was proposed for SN\,2007on is an explosion triggered by
the direct collision of two WDs, as outlined by \citet{rosswog09} and
\citet{kushnir13}. \citet{kushnir13} and following papers do not expect the two
WDs to explode independently. On the other hand, they do predict a 2-component
ejecta structure, which could give rise to some double-peaked emission lines, if
the collision occurs with a non-negligible impact parameter \citep{dong15}. 
\citet{raskin10} conducted calculations of WD explosions triggered by collisions
for different mass ratios, and found that a large range of \Nifs\ masses can be
obtained for different ratios for the two components, depending on details of
the system.  The two 1-zone components that we used to reproduce the nebular
spectrum of SN\,2007on have similar \Nifs\ mass and only slightly different
total mass, possibly suggesting also similar masses for the two white dwarfs
that may have collided. 

As we discussed above, the two emitting regions that the 1-zone models define
are located in the inner part of the SN ejecta, and their mass -- even when
combined -- should be smaller than the total mass ejected in SN\,2007on.  Taken
individually, the two components have a small mass when compared to the mass
enclosed within the same outer boundary velocity of typical Chandrasekhar-mass
models.   This may suggest that, if the two components correspond to two
different WDs, each WD may have had a mass smaller than the Chandrasekhar mass.
On the other hand, their combined mass may have been close to it. A  ``rebound''
of the two components may take place if the collision is not perfectly head-on
and the impact parameter is significant (E. Livne, private communication). If
the explosions are independent, the low masses of the two individual components
may imply low central densities of the two WDs, which would be in line with the
low abundance of neutron-rich nuclear statistical equilibrium (NSE) species
deduced for SN\,2007on. Any Ni directly observed in a SN\,Ia spectrum at such
advanced epochs must have been synthesized as stable Ni as most radioactive
\Nifs\ will have decayed by then.  Stable Fe was used in the models only to
match its expected abundance relative to that of stable Ni, but its mass is so
small that it does not really affect the thermal balance of the ejecta. Another
argument that may be in favour of the collisional scenario is the kinetic energy
of the explosion.  \citet{ashall2018} find that SN\,2007on may be compatible
with a normal explosion energy ($\KE \sim 1 - 1.3 \times 10^{51}$\,\ergs). Given
the rather low kinetic energy yield from thermonuclear burning in a SN that
produced little NSE material, having to unbind two low-mass WDs, which would
have very low binding energy, would make it easier to balance the energy
produced and that observed. 

Another possibility may be that in SN\,2007on we are witnessing an off-centre
ignition of a Chandrasekhar-mass C-O WD which then explodes  via a delayed
detonation. The total \Nifs\ mass and other properties of SN\,2007on are in line
with some of the least energetic delayed detonation models
\citep{hoeflich2002,gall2018}. The lack of stable, neutron-rich NSE species may
possibly be explained by a low central density. In the Chandrasekhar-mass
scenario, this would require a rapidly rotating WD with low binding energy. The
resulting nucleosynthesis, even if it not as effective in producing energy, may
be sufficient to unbind the WD and impart it a low \KE, such as what is seen in
SN\,2007on \citep{ashall2018}. Achieving a normal or high \KE\ would however be
difficult. The main problem for this scenario is the presence of two distinct
sets of lines, and therefore emitting regions, in the nebular spectra.  
Delayed-detonation models, even in 3 dimensions, do not seem to produce separate
components \citep[\eg][]{gamezo2005}, nor do models of the violent merger of two
sub-Chandrasekhar mass WDs \citep[\eg][]{dan2015}. Some delayed-detonation
models characterised by off-centre ignition produce significantly elongated
distributions of the Fe-group material, with some expanding at rather high
velocity relative to the centre of the WD \citep{fesen2007}. If observed under a
favourable angle, those structures might mimic the behaviour of two individual
components. On the other hand, the two nebulae that we need to reproduce the
spectrum of SN\,2007on seem to have remarkably similar mass. It may not be easy
to ``break'' the progenitor into two parts of similar mass even in a very
off-centre explosion. Additionally, the very low abundances of stable Fe in
SN\,2007on may be difficult to achieve in any Chandrasekhar-mass model. In our
opinion these are strong arguments against a single, Chandrasekhar-mass
progenitor for SN\,2007on. 

If SN\,2007on was indeed the result of the collision of two WDs, or a
very off-centre delayed detonation, are there any features other than
double-peaked nebular emission line profiles that would make it stand out from
the average SN\,Ia? This is equivalent to asking the question: Could or should
we have detected the presence of two separate cores in the early data of
SN\,2007on? The outer layers of a SN\,Ia expand at velocities much larger than
both the boundary velocities of the nebular emission lines and the separation
velocity of the two cores. It is therefore quite likely that at early times the
outer layers would just have appeared as one SN explosion \citep{ashall2018}.  
Still, if the \Nifs\ that powers the light curve is indeed located in two
distinct components, as indicated by the nebular models, the light curve might
be more like the sum of two low-mass ejecta light curves. The two separate
\Nifs\ cores would contribute to the luminosity, but the mean free path from
each component would be longer than in the case of a single, central radioactive
core. To first order, the light curve may be assumed to be the sum of those of
two individual low-mass components, and the diffusion times consequently
shorter. The light curve would rise and decline rapidly, but it would be more
luminous than the light curve of a SN with similar \Nifs\ mass concentrated in
one place. Thus, one possible signature is a high peak luminosity for the light
curve shape. This is indeed the case for SN\,2007on.

Other peculiar aspects of SN\,2007on that may be linked to an unusual
progenitor/explosion mechanism are:

Its location was outside an elliptical galaxy. Relatively few SNe\,Ia are
observed in elliptical galaxies, and they show a predominance of fast decliners 
\citep{hamuy2000,howell2001,sullivan10} and of peculiar events in general
\citep{ashall2016a}. The remote location of SN\,2007on may indicate that the
progenitor was quite an old system.

The possible detection of X-rays before the explosion \citep{vossnelemans08}.
This was later retracted on positional grounds \citep{Roelofs08}, but if the
detection was real it may have signalled pre-collision/explosion interaction
between two white dwarfs. An off-centre delayed detonation would not be expected
to show such precursors. Since the detection is quite uncertain, however, it
cannot be used as evidence for or against either scenario. 

A slowly-declining UV light curve after maximum \citep{milne10}. This may
actually be a signature of interaction not so much with a CSM, as hypothesised
in that paper, but rather of the two colliding ejecta components. This does not
mean that the ``beginning'' of the interaction was when the light curve started
to show a plateau. Interaction should have started at the time of explosion, but
its signature may only have become strong when both the densities and the
velocities of the interacting material were large enough. The details would
differ for the case of collision and off-centre ignition, but at the earliest
times the size of the interaction region would be quite small. Additionally, the
SN is naturally bright early on, and the added luminosity coming from
interaction may only have been appreciable when the SN had become intrinsically
dimmer. This is particularly true in the case of an off-centre explosion, in
which case the interaction would take place deep within the optically thick
ejecta at early times. Only one other SN in the sample of \citet{milne10},
showed a similar behaviour (actually much more pronounced). This was SN\,2005ke,
also a sub-luminous SN\,Ia. 

There may also be other features to look out for, in cases when the double-peak
signature is not sufficiently clear (a possible reason for this may be
unfavourable orientation). Some possibilities are: 

{\em Broad nebular emission lines for the light curve shape}. Most SNe\,Ia have
nebular spectra that are characterised by the same emission lines, with similar
ratios, but differing in width. More luminous SNe tend to have broader emission
lines, suggesting more complete burning, but the similarity in line ratios
indicates that the density is similar in most SNe.  Broad emission lines in SNe
that are not particularly luminous may indicate the presence of two components
separated by too small a line-of-sight velocity to be seen as separate emission
peaks or simply not well resolved by the observations. The 1-zone fit for the
entire spectrum of SN\,2007on requires a nebular velocity of 6500\,\kms, which
is typical of brighter SNe than SN\,2007on. This signature would however also
not be detected if the two cores were expanding in a direction perpendicular to
our line-of-sight.

{\em Peculiar light curves}. The presence of two nebulae containing \Nifs\ could
be compared to the presence of two SNe. The observed light curve would be some
combination of the two components. This would cause a more luminous light curve
than the light curve shape would predict. Interaction between the two ejecta
would be additional to this and may affect the UV emission, such as what was
seen in SN\,2007on, and possibly the X-ray emission as well. In the case of
SN\,2007on the UV light curve was peculiar in that while in the first 10 days
after maximum it declined more rapidly  than normal SNe\,Ia, as did the optical
light curves, after this time the decline  became slower than what is typical
for normal SNe\,Ia \citep[][figure 12]{milne10}.  Optical bands do not show
peculiarities in the case of SN\,2007on, but in other cases they might. 

%{\bf Interestingly, both SN\,2007on and SN\,2011iv show a second maximum in the
%NIR infrared band $\sim 20$ days after maximum, which is not normally seen in
%rapidly-declining SNe\,Ia.} 

Do other SNe\,Ia share the characteristics and potentially the mechanism of
SN\,2007on? A search for double peaked emission has only led to the
identification of a handful of potential candidates, SN\,2007on being the
strongest case \citep{dong15}. Of course, detection of double peaks depends on a
favourable orientation. There may be SNe that explode through the same mechanism
but do not show clearly separated nebular spectra. A possible case is discussed
below.

If the collision resulted in two components moving at a large angle with respect to the line-of-sight, we would not see double peaked emission lines. One possible signature would the be the large strength of the emission in rather narrow lines. This case, which represents the majority of the possible orientation angles, may be difficult to appreciate. In any case, for every one SN that shows double peaks there should be several that don't because of unfavourable oriantation, suggesting that WD collisions are a non-negligible fraction of at least transitional SNe\,Ia.

\section{Was SN\,2011iv like SN\,2007on?}
\label{sec:11iv}

We have shown that a 2-component model appears to be a viable solution for the
nebular spectrum of SN\,2007on. Another peculiar, transitional SN\,Ia, 
SN\,2011iv, was discovered in the same passive galaxy in the Fornax cluster and
was intensively studied both from the ground \citep{gall2018} and with the
Hubble Space Telescope \citep{foley12}. Its location is less remote than
SN\,2007on: it lies close to the effective radius of the galaxy
\citep{gall2018}. Despite being a rather rapidly declining SN\,Ia, with $\Dm =
1.79$\,mag \citep{gall2018}, which places it at the luminous end of the 1991bg
class, SN\,2011iv did not show a peculiar spectrum and had a luminosity
comparable to that of some underluminous but still spectroscopically normal
SNe\,Ia such as 2004eo, which have more slowly evolving light curves 
\citep[$\Dm \sim 1.4$--1.5;][]{pastorello04eo,mazzali04eo}. 

In a parallel paper, \citet{ashall2018} perform abundance tomography of
SN\,2011iv and find that satisfactory agreement can be reached if the SN was a
Chandrasekhar-mass, low energy ($\KE \approx 9 \times 10^{50}$\,erg) event that
produced $\approx 0.31\,\Msun$ of \Nifs\ and a relatively large amount of stable
NSE species, especially iron, similar in many ways to SN\,1986G
\citep{ashall2016b}. The nebular fit in that paper is good but not perfect. The
nebular spectrum of SN\,2011iv does not show split or double-peaked emission
line profiles, possibly because of a relatively low signal-to-noise ratio, but
it does show rather broad emission lines. The spectrum has a lower
[\FeIII]/[\FeII] peak ratio than that of SN\,2007on, indicating a lower overall
degree of ionization, similar to SN\,2004eo. SN\,2011iv satisfies at least two
of the criteria laid out above for a double-core event: a high peak luminosity
and broad emission lines for its decline rate. Thus motivated, we modelled its
nebular spectrum in order to test whether a 2-component model can yield an
improved fit. We followed the same strategy adopted for SN\,2007on above.

\begin{figure}
 \includegraphics[width=88mm]{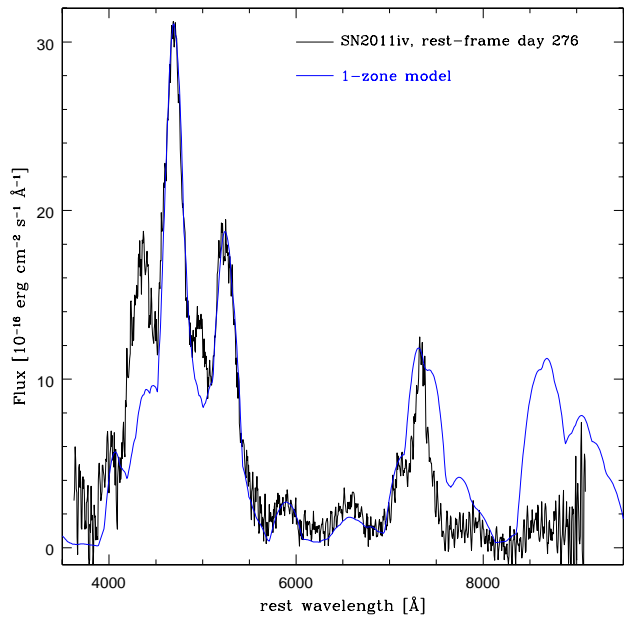}
\caption{The nebular spectrum of SN\,2011iv (black) and a 1-zone, broad-line 
model (blue).}
\label{fig:sn2011iv_1z}
\end{figure}

We modelled the spectrum obtained 260 days after $B$ maximum \citep{gall2018}.
We used a distance modulus $\mu=31.20$ as for SN\,2007on, and no reddening.
Correcting for the rotation of the host galaxy \citep{scott2014}, the recession
velocity at the position of SN\,2011iv is 1872\,\kms.  Assuming a risetime of
17.9 days, as determined by \citet{gall2018}, we used a rest-frame epoch of 276
days for the spectrum after correcting for time dilation.  

We began with a 1-zone model, which is shown in Fig. \ref{fig:sn2011iv_1z}.  An
outer boundary velocity $v = 9000$\,\kms\ is required to match the observed line
width with blends of Fe emission lines.  This is a rather high velocity
considering the decline rate of the SN. For example, SN\,2004eo, whose light
curve decline rate was slower \citep[$\Dm = 1.47$\,mag;][]{pastorello04eo}, had
a line width of $v = 7400$\,\kms. The fit to SN\,2011iv required $0.41 \Msun$ of
\Nifs. \citet{gall2018} estimate M(\Nifs) $\approx 0.42\,\Msun$ from the peak
luminosity. The total mass included within the outer boundary is $\sim
0.80\,\Msun$. Stable Fe ($0.25\,\Msun$) is the second dominant contribution
after \Nifs, followed by stable Ni ($0.08\,\Msun$). Stable Fe-group elements are
required to keep the ionization level and the ratio of the two strongest iron
emission features low.  Intermediate-mass elements also contribute, but their
masses are quite small ($0.05\,\Msun$ for Si and S combined). 

\begin{figure*}
 \includegraphics[width=134mm]{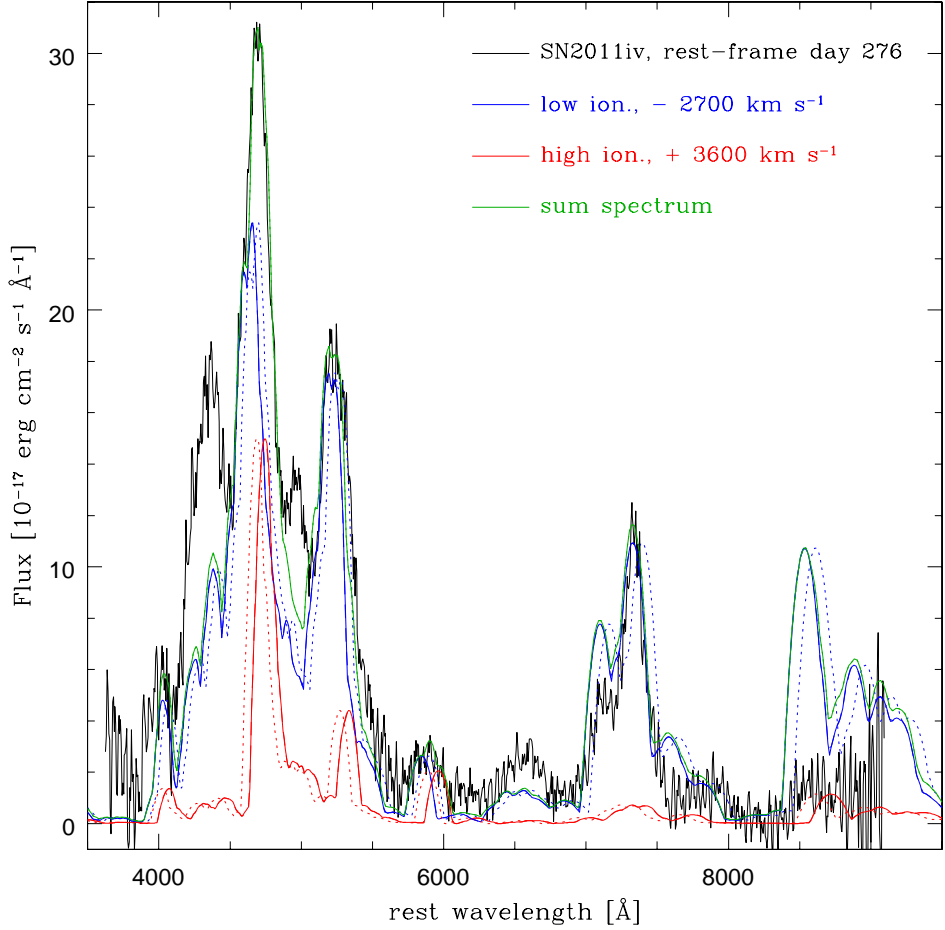}
\caption{The nebular spectrum of SN\,2011iv and the 2-component model. The dotted lines show the two components at rest.}
\label{fig:sn2011iv_2comp}
\end{figure*}

The biggest problem with building a 2-component model is that, unlike the case
of SN\,2007on, the spectrum does not immediately suggest what the shift of the
two components may be or their overall properties. One line that is strikingly
shifted with respect to its rest position is the peak near 7300\,\AA. If that
line is \NiII\,7380\,\AA, as is often assumed despite uncertainty caused by
blending with other lines such as [\CaII], then the line, which is observed at
$\approx 7315$\,\AA\ after correcting for redshift, is blue-shifted by $\sim
2700$\,\kms. This velocity could be used for one component. The strength of the
\NiII\ emission in that component means that n-rich species should have been
synthesised there. The presence of these species then implies that this
component of the spectrum will likely have rather low ionization. The other
component would then need to be more highly ionized in order to reproduce the
strong [\FeIII] emission peak near 4700\,\AA. This is a similar situation as in
SN\,2007on. The ``high-ionization'' component needs to be slightly red-shifted
in order to match the position of the observed emissions. We can optimise the
red-shift by trying to match the width of the emission peak. Our solution is
shown in Fig. \ref{fig:sn2011iv_2comp}.

We found that a good solution for the ``low-ionization'' component is given by a
nebula with boundary velocity $v = 5750$\,\kms, a total mass inside this
velocity of $0.46\,\Msun$, of which M(\Nifs)$= 0.26\,\Msun$, and a blueshift of
2700\,\kms. As in the 1-zone model, stable Fe is the second most abundant
element ($0.12\,\Msun$), but the mass of stable Ni is relatively small, $0.03
\Msun$. With the blueshift that we adopted, the [\NiII] $\lambda 7380$ line is
now at the wavelength of the observed emission, while other lines match the blue
part of the observed features, including the \NaI\,D/[\CoII] line.
Intermediate-mass elements have small abundances. Their total mass is $\sim
0.05\,\Msun$. Among these, silicon dominates with $0.03\,\Msun$. 

The mass of this ``low-ionization'' nebula within the outer boundary velocity is
quite similar to that of a normal Chandrasekhar-mass explosion. Taken on its
own, this nebula matches the basic relations that have been found for SNe\,Ia.
The \Nifs\ content matches the relation between \Nifs\ mass and light curve
decline rate \citep{phillips93,phillips99,zorro} given the decline rate of
SN\,2011iv. It also would match the relation between decline rate and Fe
emission line width \citep{zorro} given the \Nifs\ mass and the decline rate of
the SN. However, SN\,2011iv was significantly more luminous at peak than those
relations would predict: it was about as luminous as SN\,2004eo, but had a more
rapidly evolving light curve. It also shows broader lines than expected. This is
actually quite suggestive. The high peak luminosity and broad emission lines
could be due to the presence of a second component, which may affect the
post-maximum decline rate because it  evolves rapidly. The observed light curve
would then be the combination of the two components and could be more luminous
that the decline rate would imply. This requires the two components to be well
separated, which may be possible, as in the case of SN\,2007on. 

In order to complete the fit, a second component is needed. This should be of
higher ionization, as discussed above, and should contain little to no stable
nickel, since the [\NiII] emission is reproduced by the ``low-ionization''
nebula alone. This possible ``high-ionization'' component does not need to have
a very strong flux, and therefore it turns out to be quite small in mass. In our
solution it has $v = 5000$\,\kms, $M \sim 0.06\,\Msun$, and a redshift of
3600\,\kms, so that the line-of-sight velocity separation of the two components
is 6300\,\kms, larger than in SN\,2007on but compatible with the observed line
width.  The nebula is composed almost entirely of \Nifs\  $(0.06 \Msun$), and it
contains no stable Fe-goup elements. Only traces of intermediate-mass elements
are present, essentially Na and Ca ($\sim 10^{-5}-10^{-4}\,\Msun$), which are
necessary to give rise, in particular, to the red component of the
\NaI\,D/[\CoIII] feature (which is dominated by \NaI\,D). The mass of this
nebula is very small even when compared to the individual components in the
spectrum of SN\,2007on. It is actually comparable to the values of the emitting
nebulae in SNe like 1991bg or 2003hv.  Because of the much lower mass and
density, this component has a much higher ionization degree than the
blue-shifted component, as shown by the ratio of the [\FeIII] and
[\FeII]-dominated emissions. 

The combined 2-component model obtained by summing the blue- and red-shfted
components described above is shown in Fig. \ref{fig:sn2011iv_2comp}. Although
it is not very different from the single component models, whether 1-zone or
stratified \citep{ashall2018}, it does match the observed line width better and
it reproduces the shift of some lines. 

The combination of the two components yields a \Nifs\ mass of $0.32\,\Msun$,
which powers the light curve. This is not too different from what was derived
from the peak of the light curve ($0.42\,\Msun$) by \citet{gall2018} using a
semi-analytic method. The total mass included in the two nebulae (which have
slightly different boundary velocities, 5000 and 5750\,\kms) is $0.52\,\Msun$.
This is high for a standard-energy explosion, and comparable to the low-energy
SN\,1986G \citep{ashall2016a}. This value is in line with the results of
abundance tomography \citep{ashall2018}, which also indicate that SN\,2011iv had
a low explosion energy and consequently a large mass at low velocity.  The
line-of-sight separation velocity, 6300\,\kms, is comparable to the expansion
velocity of the two individual nebulae, which remain therefore at least
partially spacially separated, and could be fully separated if the viewing angle
was significantly different from the actual direction of motion. Of course both
components are embedded in the much faster outer ejecta, which at early times
may cause some light curve broadening because of diffusion. Detailed modelling
in 3 dimensions would be useful to test this.

Unlike the case of SN\,2007on, in SN\,2011iv the two components have very
different masses, which makes the collisional scenario much less likely. The
off-centre ignition scenario, on the other hand, could be viable.  The
blue-shifted, ``low-ionization'' component, which contains a significant
fraction of stable Fe-group material, may correspond to the neutron-rich
nucleosynthesis region that is predicted to occur in Chandrasekhar-mass models
\citep{iwamoto99}, and which may be identified with the ashes of the centre of
the WD. The red-shifted, ``high-ionization'' component does not contain n-rich
species, and may correspond to the off-centre component of the explosion.
Regions of the WD that are sufficiently far removed from the centre are unlikely
to be dense enough to synthesize n-rich species. In SN\,2011iv the explosion may
have started significantly off-centre, leading only to the synthesis of \Nifs.
The burning flame then propagated inwards causing the main explosion, which
converted the innermost regions of the WD to n-rich species.  The total mass
seen within the boundary expansion velocities surveyed by the nebular spectra is
high for a normal-energy Chandrasekhar-mass model, but could be in line with a
low-energy explosion, which is shown to produce the best fits to the early-time
spectra and the light curve in a companion paper \citep{ashall2018}. A high
relative abundance of n-rich NSE species is expected if the explosion proceeded
slowly, and is a common feature of both SN\,2011iv and SN\,1986G. The high bulk
velocity of the two components would also indicate a rather non-central ignition
point. The less massive off-centre component has the higher velocity, although
the values we derived are quite far from balancing momentum.

Finally, as this is possibly the first time SN\,Ia nebular spectra have been
analysed with an approach that does not assume spherical symmetry of the ejecta,
it is interesting to compare the properties of the spectra of SNe 2007on and
2011iv, in particular the shift of some of their emission lines, with their
early-time spectral properties. \citet{maeda10} suggested that there is a link
between the shift of the observed line which is identified as
[\NiII]$\,\lambda7378$ (and of another line identified as
[\FeII]$\,\lambda7155$) and the behaviour of the photospheric velocity at early
times as described by the evolution of the velocity of \SiII$\,\lambda6335$,
which is the defining line for SNe\,Ia.  The sense of the relation is that the
[\NiII] line is created in high neutron density ejecta where no \Nifs\ is made
and which corresponds to the region of the progenitor where burning occurred at
the highest densities. This may be the centre, but may be slightly off-centre
depending on where ignition first occurs. If burning starts off-centre the
region that is initially affected will start to expand and move with some
characteristic velocity, which will be reflected along the line-of-sight by an
observed shift of the [\NiII] emission line. \citet{maeda10} found that when
this line is blue-shifted the early velocity evolution is slow 
\citep[a ``Low Velocity Gradient'' (LVG) SN in the definition of][]{benetti2005}. 
Conversely, when the line is red-shifted the early velocity evolution is fast 
\citep[a ``High Velocity Gradient'' (HVG) SN in the definition of][]{benetti2005}. 
This may suggest that the explosion stretches out the progenitor more in the
direction opposite that of the region where burning begins. If SNe\,Ia do behave
generically this way off-centre delayed detonations may be a common scenario. 

For SN\,2007on, \citet{maeda10} report a \SiII\ velocity gradient of $\approx 85$ \kms d$^{-1}$, which makes it almost a HVG SN, and no shift in the emission lines. The SN appears to be an outlier in their distribution. Given the scenario that we adoped for SN\,2007on, namely the collision of two WDs, it would be natural to expect that some shift should be observed, as it is indeed for many other lines. We have found, however, that the [\NiII] line is very weak in SN\,2007on, and the emission near 7380\,\AA\ is a blend of red- and blue-shifted [\CaII], so we do not expect the relation to hold, which is indeed the case. The velocity gradient at early times is caused by the side of the ejecta that face us. If we look at the blueshfted nebula in our model, this has a velocity of 4200\,\kms, which is extreme when looking at Figure 2 of \citet{maeda10}. Therefore, even if we only consider the blue-shifted component of the nebular spectrum, SN\,2007on does not fit in the relation of \citet{maeda10}. This may be seen as evidence supporting a different explosion scenario for SN\,2007on. 

As for SN\,2011iv, we suggest here that this is likely to be an off-centre delayed detonation. It should therefore be the ideal candidate to test the scenario of \citet{maeda10}. \citet{ashall2018} find for SN\,2011iv a \SiII\ velocity gradient of $120 \pm 20$ \kms d$^{-1}$, which places it squarely among the HVG events. These are thought to be show red-shifted [\NiII] emission lines. The [\NiII] line in SN\,2011iv, however, is actually blue-shifted by $\sim 2700$\,\kms. Again, this SN does not seem to follow the relation suggested by \citet{maeda10}.
While this does not mean that SN\,2011iv could not be an off-centre delayed detonation, it may indicate that the relation of \citet{maeda10} is not always predictive. What may actually be off-centre is not the ignition point but rather the point of transition from a deflagration to a detonation \citep{fesen2007}.

In conclusion, SN\,2007on appears to be consistent with the explosion following
the collision of two WDs. An alternative - but in our opinion less likely -
scenario is a very off-centre delayed detonation of a Chandrasekhar mass WD. 
This latter scenario is more likely to apply to SN\,2011iv. It is quite likely
that most, if not all, physically motivated mechanisms that have been proposed
do actually exist. Both direct collisions and off-centre delayed detonations may
be more common among fast decliners. Our task is then to identify possible
observed counterparts for these mechanisms. Being able to distinguish between
different progenitor/explosion channels would greatly help with regards to using
SNe\,Ia as cosmological standardizable candles. Nebular spectroscopy is a very
useful tool to investigate the inner workings of SNe\,Ia.

\section*{ACKNOWLEDGEMENTS}

We thank Wolfgang Hillebrandt and Michele Sasdelli for useful conversations.
The Carnegie Supernova Project (CSP) is supported by the National Science 
Foundation (NSF) under grants AST-20130306969, AST-20130607438, 
AST-20131008343, AST1613426, AST1613455 and AST1613472.
P.A. Mazzali and C. Ashall acknowledge support from STFC.
Supernova research at Aarhus University is supported in part by a Sapere Aude 
Level 2 grant funded by the Danish Agency for Science and Technology and 
Innovation, and the Instrument Center for Danish Astrophysics (IDA). 
M. Stritzinger is also supported by a research grant (13261) from VILLUM FONDEN.
C. Gall acknowledges support from the Carlsberg Foundation. 
P. Hoeflich acknowledges support by the National Science Foundation (NSF) grant 1715133.
Finally, we wish to thank the ananymous referee for a constructive report.

\bibliographystyle{mn2e}

\end{document}